# Thermal Stability of Diamond-Like Carbon Nanothreads

L.A. Openov, A.I. Podlivaev

National Research Nuclear University "MEPhI", Kashirskoe sh. 31, Moscow 115409, Russian Federation

E-mail addresses: LAOpenov@mephi.ru, laopenov@gmail.com

**ABSTRACT**

The thermally activated fracture processes in the carbon backbone of diamond-like carbon nanothreads and the hydrogen desorption from them has been studied by the molecular dynamics method. Specifically, the temperature dependence of the characteristic desorption time at $T$ = 1700−2800 K has been determined. The activation energy and frequency factor in the Arrhenius formula for the desorption rate are found. This allows estimating the desorption time at any temperature. The mechanical stiffness of nanothreads is calculated.

After the discovery of graphene [1], many other quasi-two-dimensional carbon-based materials have been predicted [2–4] and synthesized [5, 6], including T-graphene, octagraphene, phagraphene, graphyne, graphdiyne, etc. The combination of carbon with hydrogen largely extends this list by adding to it graphane (graphene hydrogenated from both sides) [7,8], one-side hydrogenated graphene [9, 10], graphone [11], diamane [12], etc. In the quasi-one-dimensional nanostructures such as carbon nanotubes (CNTs) [13], hydrogen also plays an important role, often drastically modifying their thermal and electrical characteristics. In particular, after saturation of the metastable metallic (3, 0) and (2, 2) CNTs with hydrogen up to the stoichiometric composition C:H = 1:1, they become stable and insulating [14].

Recently discovered diamond nanothreads [15] are obtained from the (3, 0)H CNT by a series of Stone–Wales (SW) transformations [16], i.e., by the rotations of C–C bonds (with the adsorbed hydrogen atoms) at an angle of about $90^0$. Here, large regions with the (3, 0)H CNT structure are separated by compact areas with two SW transformations in each of them (Fig. 1). These areas are randomly located along the nanothread. They are often referred to as SW defects [17, 18] using the terminology accepted for the defects arising in graphene owing to the SW transformation [19]. However, we should keep in mind that while such areas for (3, 0)H CNTs are indeed related to defects (breaking the translational symmetry), they are inherent in the structure of nanothreads because the rotated C–C bonds "dam up" the CNT cage, giving rise to the formation of unique objects differing from both CNTs and usual polymers. In nanothreads, all valence carbon orbitals are $sp^3$-hybridized, in much the same way as in diamonds, in contrast to single-wall CNTs with the $sp^2$ hybridization of carbon orbitals.

In the applications involving strong heating of the sample, it is necessary to know the borders of the stability range for the nanothreads in the temperature–time plane. The aim of this work is to determine the possible channels and characteristic times for the fracture of nanothreads heated to high temperatures using molecular dynamics (MD) simulations of their evolution. We demonstrate that the main reason for the decay of the characteristic structure of the nanothreads is thermally activated hydrogen desorption, whereas the integrity of the carbon backbone is retained on heating with rare exception.

For the simulation of a nanothread, we used a fragment of the (3, 0)H CNT consisting of 240 atoms and containing one SW defect (i.e., two C–C bonds rotated by an angle of about 90°, see Fig. 2). We used periodic boundary conditions along the CNT direction and free ones in both transverse directions.

The interatomic C–C, C–H, and H–H interactions are described in the framework of the nonorthogonal tight-binding model [20], which is less accurate than ab initio methods but demands much less computer resources and provides an opportunity to study the evolution of systems consisting of several hundred atoms in detail for a quite lengthy period of time (at the atomic scale). In contrast to the classical approaches, this model explicitly takes into account the quantum contribution to the total energy related to the electron subsystem. For carbon and hydrocarbon nanostructures, it provides the binding energies and interatomic distances that agree well with the experimental data and ab initio calculations (see [20–22] and references therein). Earlier, we used this model in the simulations of the thermal desorption of hydrogen from graphane [21], interactions of SW defects in graphene [23], etc.

After the detailed study of the potential energy surface (PES), we found that the atomic configuration shown in Fig. 2 is metastable; i.e., it corresponds to a local minimum in PES (there are no imaginary frequencies in the spectrum of vibrational eigenmodes). The energy of this configuration is 1.1 eV higher than that of the initial (3, 0)H CNT. The height of the energy barrier preventing the formation of our model nanothread from (3, 0)H CNT and corresponding to the saddle point in the PES is 4.2 eV, whereas the height of the barrier to the annealing of the nanothread (i.e., to its return to (3, 0)H CNT owing to two inverse SW transformations) is $U_a = 3.1$ eV.

The simulations of the time evolution of the nanothreads is performed by the MD technique (see details in [21]). For the numerical integration of the equations of motion, we use the Verlet algorithm. We record the positions of all atoms every ten MD steps. Then, the data obtained are represented in the form of a computer animation. This provides visual information on the character of distortions in the carbon backbone and allows us to fix the time of the desorption of a hydrogen atom or molecule with a good accuracy.

The calculations are performed for $T = 1700-2800$ K. Such a high value for the lower boundary of this range is chosen because the rate of any thermally activated process decreases exponentially upon lowering the temperature according to the Arrhenius law

$$\tau^{-1}(T) = A \exp\left[-\frac{E_a}{k_B T}\right], \quad (1)$$

where $\tau$ is the characteristic time of this process, $E_a$ is its activation energy, $A$ is the frequency factor in units of inverse seconds, and $k_B$ is the Boltzmann constant. As temperature decreases, this leads to the exponential growth of $\tau$ and to the corresponding growth of the computer time needed for the calculations. In our work, we use an alternative approach [21] using the direct numerical determination of $\tau$ at high temperatures and the extrapolation of the results to the ranges of intermediate and relatively low temperatures.

In the overwhelming majority of the cases under study (about 90%), the fracture of $C_{120}H_{120}$ nanothread (Fig. 2) starts with the separation of one hydrogen atom ($\approx 70\%$ of cases) or $H_2$ molecule ($\approx 20\%$) from it. The separation of the first hydrogen atom (first molecule) from the (3, 0)H CNT region occurs about an order of magnitude more often than that from the defect region. This corresponds to the ratio of the sizes of these regions and of the number of hydrogen atoms in them, respectively. We treat the lifetime of the heated nanothread up to the time of the first separation event as the characteristic desorption time $\tau$ (for the nanothreads of an arbitrary size, this time corresponds to the desorption of about 1−2% of hydrogen from it).

The results of our "computer experiment" are illustrated in Fig. 3. We can see that, in a wide (exceeding two orders of magnitude) range of $\tau$, the dependence of the logarithm of $\tau$ on the inverse temperature is well fitted by a straight line according to the Arrhenius law (1). The slope of this line gives the activation energy in relation (1), whereas its intersection with the vertical axis determines the frequency factor. Taking into account a relatively large scatter of the data (owing to the probabilistic nature of the desorption process), we determined $E_a = (2.9 \pm 0.2)$ eV and $A = (4.6 \pm 0.6) \times 10^{17}$ s$^{-1}$. Using these values of $E_a$ and $A$, we can estimate, according to Eq. (1), the desorption time at temperatures lying outside the range at which the calculations are performed. At $T = 300$ K, the desorption time is macroscopically large ($\tau > 10^{30}$ s). Even on heating to $T = 400°C$, the value of $\tau$ remains very large (~ 3 h). The further increase in the temperature leads to a fast decrease of $\tau$ to about 10 s at $T = 500°C$ and to 0.1 μs at $T = 1000°C$.

In four cases, the hydrogen desorption was preceded by the fracture of the carbon backbone of the nanothread. Three times, it began from breaking of the C–C bond in the (3, 0)H CNT region, and once in the defect region, propagating over the whole nanothread owing to the subsequent breakings of new and new C–C bonds. The broken bonds never recover; i.e., the fracture of the carbon backbone becomes irreversible as soon as it starts. Insufficient statistics impedes the determination of the temperature dependence of the characteristic decay time for the backbone fracture. We can only say that this time exceeds as a rule the time of the beginning of the hydrogen desorption; i.e., the desorption is the main channel for disturbing the nanothreads structure, at least, at high temperatures.

We have never observed the annealing of the defect region in the nanothread, i.e., two inverse SW transformations. This is because, first, the annealing activation energy (which in the first approximation is equal to the height $U_a$ of the barrier for annealing) exceeds the activation energy for hydrogen desorption and, second, the frequency factor for annealing, $A_a = 9 \cdot 10^{14}$ s$^{-1}$ (calculated according to the Vineyard formula [24]), is smaller than that for desorption. As a result, it follows from Eq. (1) that the hydrogen desorption at any temperature begins much earlier than the annealing of the SW defect takes place.

In conclusion, let us give some comments on the effect of hydrogen desorption on the elastic characteristics of nanothreads. For quasi-one-dimensional systems, the notion of the area of a cross section is meaningless. Therefore, following [17], we define the nanothreads stiffness $E$ as the coefficient of proportionality between the tensile force and relative elongation $\varepsilon$. At $\varepsilon \ll 1$, we find $E = 135$ nN. This value is much lower than the stiffness $E = 197$ nN of the initial (3, 0)H CNT; i.e., the defect region "softens" the nanotube. After the thermally activated desorption of a hydrogen atom or molecule, the stiffness remains nearly the same, being within the range of 135–140 nN. Assuming that the effective diameter of the nanothread equals 5 Å [17], we obtain $E = 689$ GPa (in customary units). This value is close to the Young modulus for single-wall non-hydrogenated CNTs [25].

Thus, the diamond-like carbon nanothreads are characterized by fairly high thermal stability: their specific structure does not change for a long time on heating to 300–400°C, and their mechanical strength is retained even after desorption of 1–2% of hydrogen atoms. The obtained temperature dependence of the nanothread lifetime should be taken into account in the choice of the fields of their possible applications, especially those dealing with high working temperatures.

This work was performed at the National Research Nuclear University MEPhI and was supported by the Russian Scientific Foundation, project no. 14-22-00098.

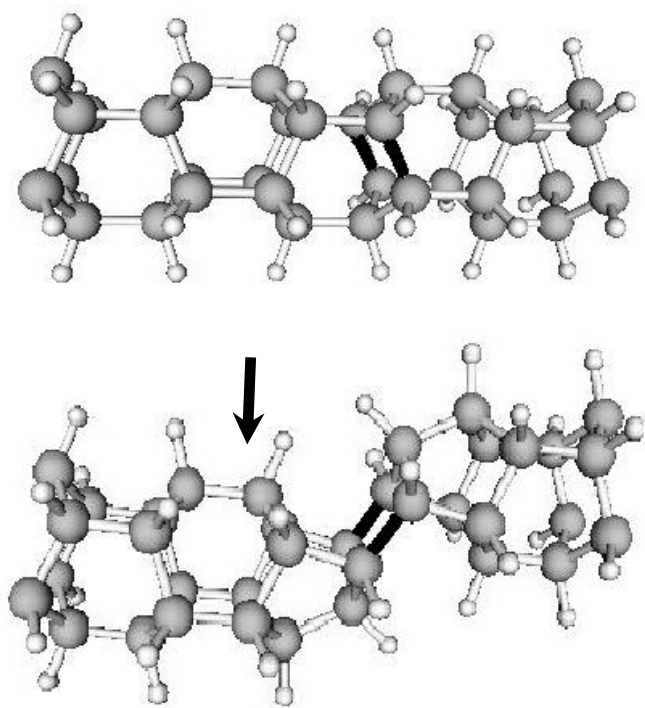

**Fig. 1.** Schematic picture of the formation of the defect portion in the nanothread. Large and small spheres denote carbon and hydrogen atoms, respectively. The C–C bonds rotating by an angle of about $90^0$ in the course of SW transformation are highlighted in black.

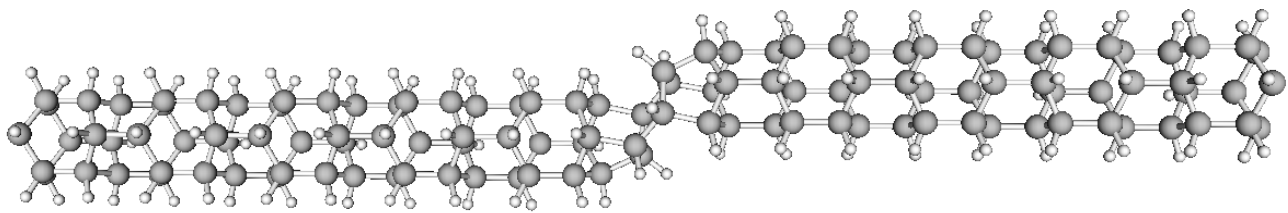

**Fig. 2.** $C_{120}H_{120}$ nanothread with one defect region

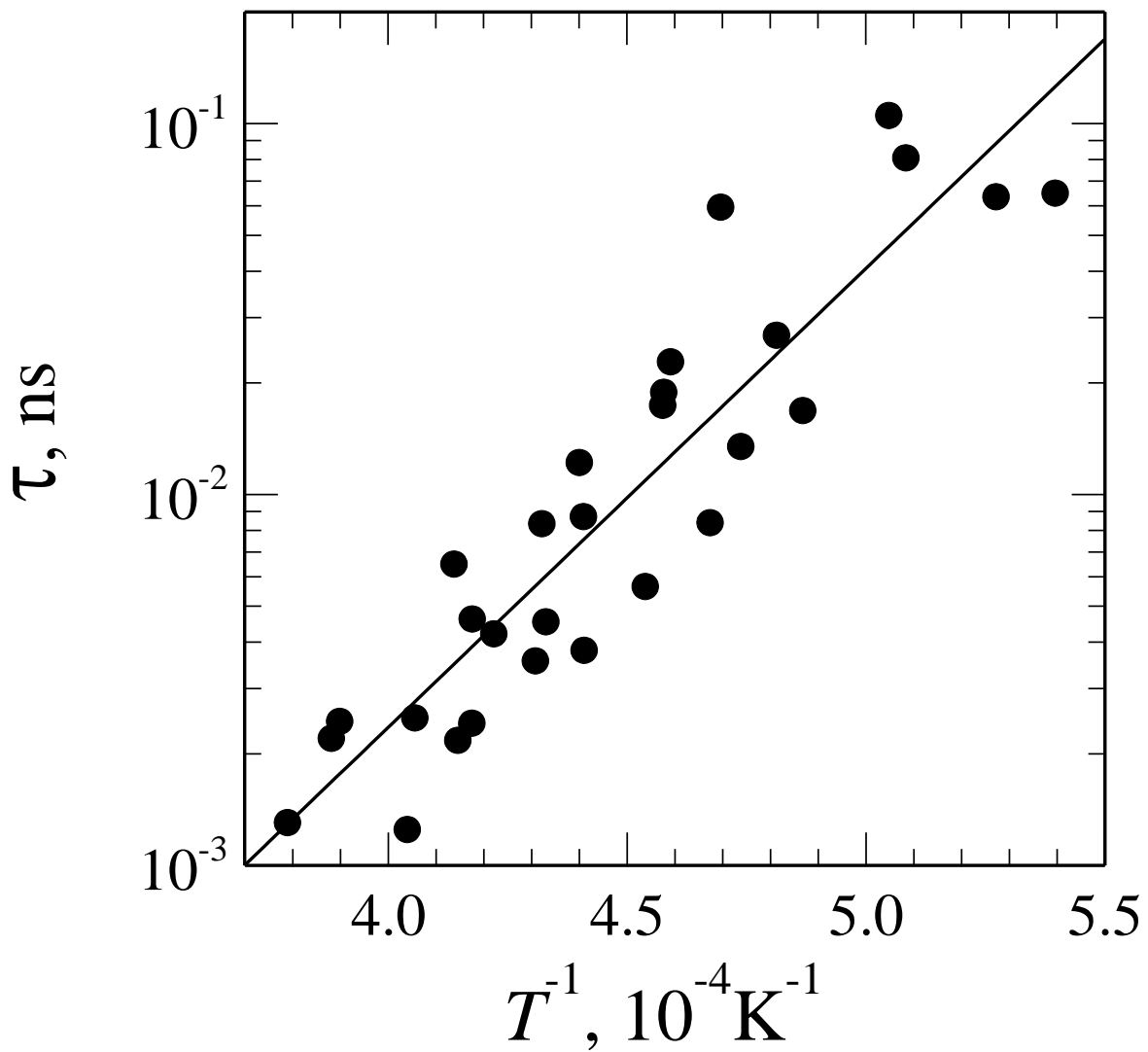

**Fig. 3.** Time characterizing the desorption of one hydrogen atom or molecule from the $C_{120}H_{120}$ nanothread (see Fig. 2) versus the inverse temperature $T^{-1}$. Points correspond to the results of numerical calculations and the solid line is the linear least-squares fit.